\documentclass[10pt]{iopart}

\usepackage{iopams}
\usepackage{graphicx}
\usepackage{bm}
\usepackage{cite}
\usepackage{mathrsfs}

\newcommand{\text}[1]{\mathrm{#1}}
\newcommand{\eqref}[1]{(\ref{#1})}

\newcommand{\const}{\ensuremath{\text{const}}}
\newcommand{\abs}[1]{\ensuremath{\left| #1\right|}}
\newcommand{\eps}{\ensuremath{\varepsilon}}

\newcommand{\be}{\begin{equation}}
\newcommand{\ee}{\end{equation}}

\newcommand{\Ga}{\ensuremath{\Gamma}}

\newcommand{\f}[1]{\ensuremath{\boldsymbol{#1}}}

\newcommand{\df}{\ensuremath{\mathrm{d}}}

\renewcommand{\etal}{ \emph{et\,al.}}

\begin{document}

\title[Isolated unstable Weibel modes]
{Isolated unstable Weibel modes in unmagnetized plasmas with tunable asymmetry}

\author{R.\,C. Tautz$^1$ and I. Lerche$^2$}
\address{$^1$Institut f\"ur Theoretische Physik, Lehrstuhl IV: Weltraum- und Astrophysik,
Ruhr-Universit\"at Bochum, D-44780 Bochum, Germany\\
$^2$Institut f\"ur Geowissenschaften, Naturwissenschaftliche Fakult\"at III,
Martin-Luther-Universit\"at Halle, D-06099 Halle, Germany}

\eads{$^1$\mailto{rct@tp4.rub.de}, $^2$\mailto{lercheian@yahoo.com}}

\date{\today}

\begin{abstract}
In this paper, an initially unmagnetized pair plasma with asymmetric velocity distributions is investigated where any unstable Weibel mode must be isolated, with discrete values for the growth rates and the unstable wavenumbers. For both a non-relativistic distribution with thermal spread and a high-relativistic two-stream distribution it is shown that isolated modes are excited and that, as the asymmetry tends to zero, the growth rate remains finite, as long as the distribution function is not precisely symmetric.\end{abstract}


\section{Introduction}

Ever since the classic papers about the Weibel instability \cite{wei59,fri59} showing the presence of kinetic instabilities in an infinitely extended homogeneous plasma with the instability dependent solely on bulk properties of the plasma and independent of resonant wave-particle effects, there has been, and continues to be, significant interest in such exponentially growing modes that have no propagating component. Much effort has focused on electron beams passing through a background plasma \cite{ng06,bre05b}, on non-linear aspects \cite{rom04,lee73}, on the saturation mechanisms of the instability\cite{sil03,kat05} as well as on mode coupling effects \cite{tzo06,tag72,TLS06}.

One might argue that the original discussion by Weibel concerning the modes that now bear his name is limited in that one can find a frame of reference in which the modes propagate; however, in such a frame the particle distribution functions would then have bulk speeds. In the one preferred frame where there is no bulk speed of the particles the original modes discovered by Weibel are purely growing. Indeed, in a frame in which the waves would propagate, they would also have a spatial component that grows exponentially. Such spatial growth is to be avoided if possible in dealing with an homogeneous plasma because it leads to difficulties of interpretation for energy and momentum considerations. The current investigation is restricted to such reference frames [the condition for which is stated in Eq.~\eqref{eq:cond}]. Furthermore, the investigation is, strictly speaking, limited to infinitely large systems. However, with the approximation that the system size is much larger than the relevant wavelengths, it becomes applicable to real systems.

In the literature, this difference is known and, usuallly, such different types of instabilities are noted as ``absolute'' and ``convective'' instabilities.

The behavior of symmetric distributions with a temperature anisotropy \cite{rs:cov} or interpenetrating streams \cite{cal06,TS05,TS06}, which shows the presence of a broad range of unstable wavenumbers, is well understood. Various astrophysical applications are known that range from the creation of cosmological magnetic seed fields (e.\,g., \cite{sch05}) through highly relativistic processes such as the jets of active galactic nuclei and gamma-ray bursts \cite{die06} to local phenomena in the solar system. Even a photon gas can be subject to filamentational instabilities \cite{shu04}. The behavior of asymmetric distributions, however, is not as well investigated. Recently, some work has focused on the subject of unstable isolated Weibel modes with discrete wavenumbers that occur in asymmetric plasma distributions \cite{usr06a,usr06b}.

Based on the relativistic linear dispersion tensor for an initially unmagnetized plasma \cite{ler69}, recently \cite{TL06a,TL06b} (hereafter referred to as T06a and T06b, respectively) an analytical proof was given that any unstable Weibel mode \emph{must} be isolated, i.\,e., restricted to discrete wavenumber values at which the instability is excited, if one of the following conditions is fulfilled: (i) a totally asymmetric distribution function (where ``total'' refers to the fact that the distribution is asymmetric in all three spatial directions), and (ii) a gyrotropic (i.\,e., symmetric around a featured axis) distribution, in which case the angle between the axis of wave propagation and the symmetry axis of the distribution has to be oblique. The fact that there may be a frame of reference in which the modes appear as propagating is not relevant for one can always find a frame of reference in which the modes are purely growing for an asymmetric plasma. Those modes where no propagation occurs relative to the plasma, be it symmetric or asymmetric, cause an in-place growth of an instability that does not propagate away from its location in the plasma and so represents a large amplitude non-propagating aperiodic instability.
%
According to the results of T06a, isolated modes can occur for any kind of distribution that is asymmetric \emph{with respect to the axis of wave propagation}. Such naturally includes all kinds of counterstreaming distribution functions where: (i) the distribution is gyrotropic with respect to the axis of the counterstream; but (ii) the counterstream itself is asymmetric, i.\,e., the two streaming components have different relative intensities and/or different temperatures; and (iii) the axis of wave propagation has an oblique angle to the streaming direction. In fact, slight asymmetries are far more likely than ``classical'' counterstream scenarios, where the two streams are often taken to be symmetric and where the analysis is confined to strictly parallel or perpendicular wave propagation.

The isolated unstable modes that occur in linear kinetic theory are reminiscent of soliton-like waves \cite{sch75} that are also based on discrete wavenumbers, when taking into account the non-linear aspects. The radiation processes of particles scattered in soliton waves are currently under active investigation. They should provide a possibility for the observation not only in astrophysical plasmas but also and especially in machine plasmas. As a prominent example, solitons are, on Earth, important for information transport in non-linear media \cite{agr06}. In astrophysical environments, there are both theoretical investigations \cite{sta03,ler02} and observational evidence \cite{shu03} for soliton plasma waves, which might act as a tool for probing the free energy content and the streaming behavior of the plasma. In this paper, isolated unstable Weibel modes are investigated for astrophysical plasmas, which modes act as a precursor for soliton waves when taking into account the non-linear aspects of the scenario. However, after having formulated the general proof that any total asymmetric distribution leads to unstable isolated Weibel modes, it was not clear whether, for realistic two-streams and counterstreams with thermal spread, the isolated Weibel modes can be actually found. Here, this open question is addressed for the first time.

We are, of course, also aware that the isolated modes have in their neighborhoods other, weakly propagating unstable modes, which point will be investigated in great detail in a future paper. Thus, the need arises for an analytical description of non-relativistic thermal counterstreaming plasmas, which, in machine plasmas, is a much more realistic case than the ``cold'' two-stream scenario, especially because the latter requires highly relativistic streaming velocities for the occurrence of isolated Weibel modes. Furthermore, thermal effects may also play a role in astrophysical plasma that are streaming with non-relativistic velocities, like, e.\,g., the solar wind.

The importance of this work lies in the fact that \emph{precisely} symmetric plasmas are difficult to achieve in Nature so that the isolated Weibel modes will be ubiquitous. Because of their rapid growth rates (of the order of the plasma frequency) for almost all asymmetric plasma situations, such isolated Weibel modes are likely to have a dominant role in shaping the evolutionary characteristics of many astrophysical, asymmetric plasma, problems. A recent self-consistent particle-in-cell (PIC) simulation \cite{TSL07} (hereafter referred to as T07) has given some evidence that in non-relativistic thermal (i.\,e., Maxwellian) plasmas, isolated unstable Weibel modes can indeed occur. This PIC simulation underlines the relevance of the subject and gives rise to the need of an analytical description which, up to now, has only been formulated for the case of cold counterstreams, thus requiring highly relativistic streaming velocities. However, because the mathematics involved is highly tedious, here we use a generalized Cauchy distribution instead, which has the advantage that all involved integrals can be calculated analytically. As will be shown in the next section, the generalized Cauchy distribution qualitatively agrees with a  Maxwellian distribution and, therefore, provides a good and more tractable description of thermal plasmas.

Therefore, in this paper first a non-relativistic counterstreaming distribution function is investigated similar to that of T07 and, following the procedure of T06a and T06b \cite{F1}, it is shown that indeed isolated Weibel modes are excited. Next, a highly relativistic two-stream distribution is investigated which, in contrast to T06b, uses neutral streams, each of which consisting of both electrons and positrons. The question how such instabilities behave if the asymmetry tends to zero will be addressed here. Both distribution functions are constructed so that the asymmetry is tunable in order to allow for a detailed investigation of the transition to symmetry. This paper shows that, in most cases, the growth rate of the isolated modes remains effectively unchanged as the asymmetry tends to zero. Even when the asymmetry is as small as the accuracy of a standard computer algebra system allows, the asymmetric plasma still supports isolated Weibel modes. Only for precisely symmetric plasmas does the isolated mode vanish, as is also clear from the analytical investigation, which the numerical calculations support to the limit of accuracy obtainable.

\section{Non-relativistic investigation}

First, an asymmetric Cauchy distribution will be investigated that is non-relativistic and allowes for thermal spread. Starting from the one-dimensional Cauchy distribution that is given by $[1+(ap+b)^2]^{-1}$, a generalized three-dimensional Cauchy distribution is constructed. Incorporating asymmetry with respect to all spatial directions and a counterstream with different relative intensities, this distribution has the form
\be\label{eq:Cauchy}
\frac{F(\f p)}{C}=\eps\left[1+\sum_{i=1}^3\left(qap_i+b\right)^2\right]^{-3}+\eps'\left[1+\sum_{i=1}^3\left(ap_i-b\right)^2\right]^{-3}\hspace{-2.5ex},
\ee
where the sums extend over all momentum components. For simplicity, the parameters $a$ and $b$ describing the thermal and streaming velocities are assumed to be equal in all directions. The factor $C^{-1}=\pi^2/(4a^3)\eps'(1+q)$ is a normalization factor to ensure that $\int F(\f p)\,\df^3p=1$. Furthermore, the parameters $\eps\in\left]0,1\right[$ and $\eps'=1-\eps$ describe the relative intensities. In order to assure vanishing bulk velocity and vanishing particle flux in all three spatial directions (corresponding to vanishing net current), i.\,e.,
\be\label{eq:cond}
\int\df^3p\;\frac{\f p}{\gamma}\,F(\f p)\simeq\int\df^3p\,\f pF(\f p)\stackrel{!}{=}0,
\ee
one must demand that $q\simeq\sqrt[4]{\eps/\eps'}$. The approximation in Eq.~\eqref{eq:cond} is valid in the non-relativistic regime where $F(\f p)\rightarrow0$ for $\abs{p_i}\gtrsim1$. Note that the exponent at the curly brackets in Eq.~\eqref{eq:Cauchy} must be greater or equal to 3 because, otherwise, $F$ cannot be normalized. To allow for the calculation of the first and second moments of the distribution function, the exponent would have to be greater or equal to 5.

\begin{figure}
\begin{center}
\includegraphics[width=95mm]{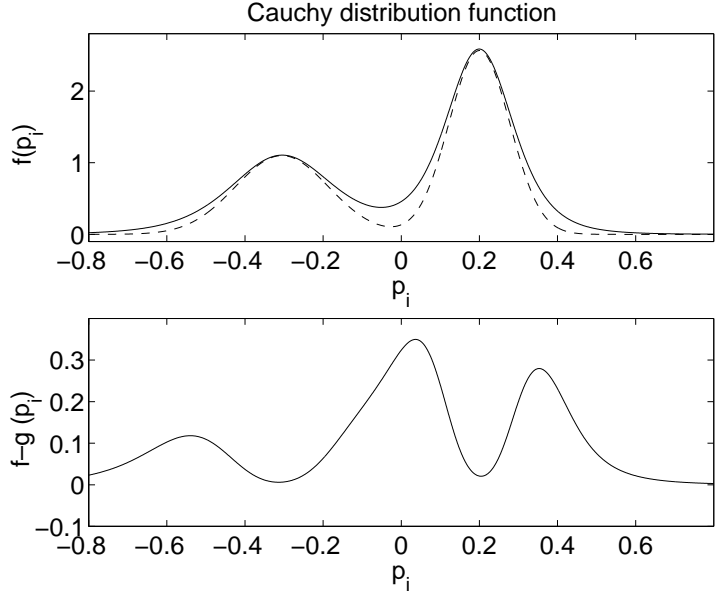}
\end{center}
\caption{The counterstreaming Cauchy distribution function from Eq.~\eqref{eq:Cauchy} for the parameters $a=5$, $b=1$, and $\eps=0.3$ (solid line) in comparison to a Maxwellian counterstreaming distribution function (dashed line) from Eq.~\eqref{eq:Maxw}. For larger values for $a$ and $b$, the difference between the two distribution functions is negligible.}
\label{ab:dist}
\end{figure}

The distribution function from Eq.~\eqref{eq:Cauchy} describes an asymmetric counterstream in all three spatial directions (see Fig.~1 in T06a) and is, therefore, comparable to that of the Maxwellian counterstreaming distribution of the form
\be\label{eq:Maxw}
\frac{G(\f p)}{C}=\eps\prod_{i=1}^3e^{-\left(qp_i+b/a\right)^2/p_{\text{th}}^2}+\eps'\prod_{i=1}^3e^{-\left(p_i-b/a\right)^2/p_{\text{th}}^2},
\ee
where again $q\simeq\sqrt[4]{\eps/\eps'}$ in the non-relativistic limit. The normalization constant is now given by $C^{-1}=\pi^{-3/2}p_{\text{th}}^3\eps'(1+q)$, where $p_{\text{th}}$
describes the ``thermal momentum.'' In Fig.~1, a one-dimensional version of the distribution function from Eq.~\eqref{eq:Cauchy} is shown in comparison to that from Eq.~\eqref{eq:Maxw}.

\begin{figure}
\begin{center}
\includegraphics[width=95mm]{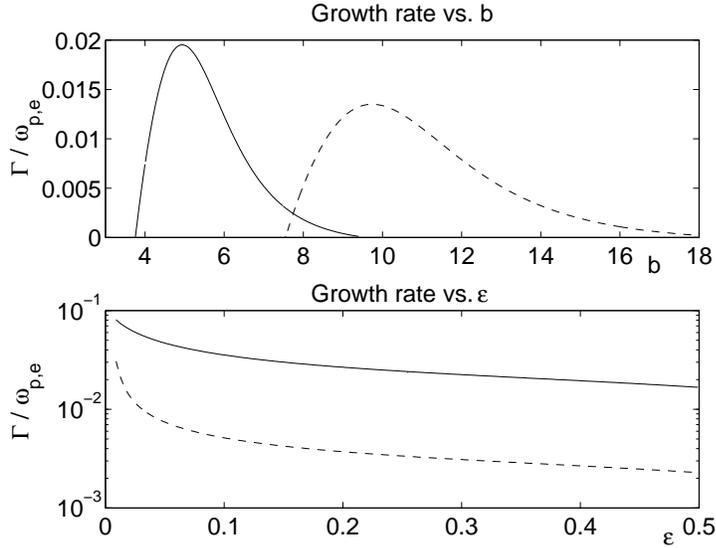}
\end{center}
\caption{The normalized growth rate $\Ga/\omega_{p,e}$ for the distribution from Eq.~\eqref{eq:Cauchy} for $a=10$ (solid line) and $a=20$ (dashed line, exaggerated by a factor $5$). The upper panel shows the growth rate for fixed $\eps=0.4$. The lower panel shows the growth rate for fixed $b=5$ (solid line) and $b=10$ (dashed line). Note that the growth rate remains approximately constant as the asymmetry tends to, but is not precisely, zero.}
\end{figure}

A distribution of the form from Eq.~\eqref{eq:Maxw} was also used in T07 for a PIC simulation, where evidence was found that, for such distributions, isolated Weibel modes are indeed excited. Note that, in contrast to the distribution function from Eqs.~(3a) and (3b) in T07, the two counterstreaming components of the distribution functions discussed here are \emph{not} symmetrically centered; however, condition~\eqref{eq:cond} still holds. The plasma is chosen to consist of an equal number density of electrons and positrons with the same distribution function, but an electron counterstream with a fixed ion background would only decrease the growth rates by a factor of $1/\!\sqrt{2}$.

Because the investigation is restricted to the non-relativistic regime, care must be taken that $F(\f p)\rightarrow0$ for $\abs{p_i}\gtrsim1$, which means that $v_i\ll c$. In that case, the Lorentz factor $\gamma=(1+\f p^2)^{1/2}\simeq1$ can be neglected in the integrals representing the dispersion relation [Eqs.~(13a) to (13e) and Eqs.~(4a) to (4e) in T06a]. Because of the lengthy and tedious algebra, we refer to T06a and T06b for further details of the derivations.

In the non-relativistic limit, all integrals representing the dispersion relation [Eqs.~(13a) to (13e) and Eqs.~(4a) to (4e) in T06a], can be solved analytically to obtain an equation for the imaginary phase velocity $M$. This implicit equation can be solved numerically, yielding real and positive values both for $M$ and the squared normalized wavenumber $\kappa^2$. As shown in T06a, a growth rate $\Ga$, normalized to $\omega_{p,e}=\sqrt{4\pi n_eq_e^2/m_e}$ (the electron plasma frequency), is given by $\Ga/\omega_{p,e}=\sqrt2\kappa M$. The growth rate as a function of the parameter $b$ (which describes the counterstream velocity) is shown in the upper panel of Fig.~2. This means that, for increasing streaming velocity (described by the parameter $b$), the growth rate has a maximum and does \emph{not} increase infinitely. From the observation of soliton waves one could, therefore, draw conclusions about the streaming velocities or, more general, the amount of free energy available in the system. In the lower panel of Fig.~2, the growth rate is shown as a function of the asymmetry parameter $\eps$. Clearly, the growth rate does \emph{not} tend to zero as the asymmetry becomes small. However, as shown in T06a, isolated Weibel modes can exist as long as the asymmetry is not \emph{precisely} zero. In that case, the equation determining $\kappa^2$ has no well-defined solution.

\begin{figure}
\begin{center}
\includegraphics[width=95mm]{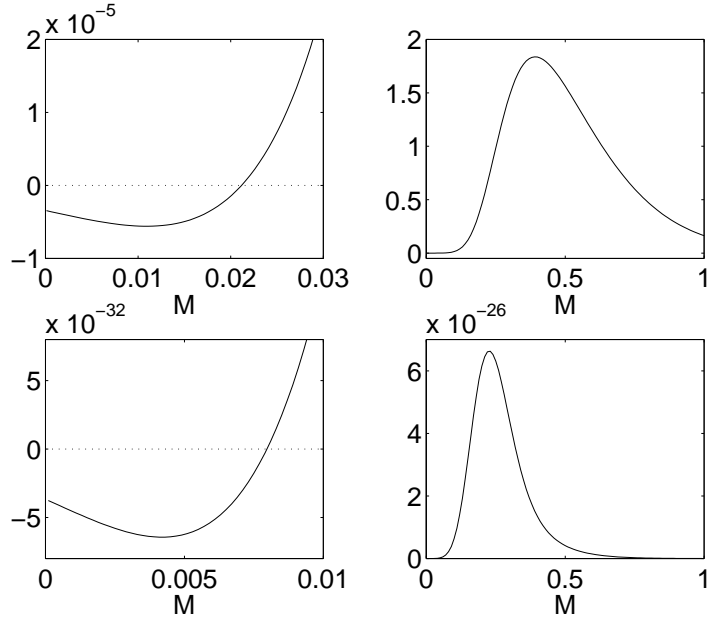}
\end{center}
\caption{The function that determines the allowed (discrete) values for $M$, i.\,e., the left-hand side of Eq.~(31) in T06b. Parameters are taken as $a=10$, $b=5$, and $\eps=0.4$ (upper panels) and $\eps=0.499$ (lower panels). For $\eps\rightarrow0.5$ (symmetric distribution), the function values become overall smaller (note the smaller scales in the lower panels) until, eventually, the function is equal to zero everywhere.}
\end{figure}

For the case $a=10$, $b=5$, and $\eps=0.4$ the function that determines the allowed values for $M$ through its zeros [i.\,e., the left-hand side of Eq.~(31) in T06b] is shown in Fig.~3. However, only one of the zeros corresponds to positive value of $\kappa^2$, therefore describing a real and positive growth rate. As the asymmetry tends to zero, the function values become smaller until they eventually reach zero, while the number of possible $M$ values remains unchanged unless the asymmetry vanishes. Therefore, the equations allow for a continuum of $M$ values \emph{only} if the distribution is exactly symmetric and the transition from the discrete unstable wavenumbers to a continuum is not smooth (see also the remarks in T06b).

\section{Relativistic investigation}

For completeness, now a relativistic distribution function is investigated that is composed of two identical electron-positron beams. In contrast to the distribution used in T06b, here the two beams are charge neutral and, more importantly, the asymmetry is tunable through a parameter $\eps$. Using the normalized momentum $\f p=\f P/(m_ec)$ (with $\f P$ the real momentum), a relativistic two-stream distribution function with tunable asymmetry is constructed as
\be\label{eq:beam}
H(\f p)=\eps\prod_{i=1}^3\delta\!\left(p_i-\varpi_\ell\right)+\eps'\prod_{i=1}^3\delta\!\left(p_i+\varpi_r\right)
\ee
where now $\varpi_r=\eps\varpi_\ell\,\bigl/\!\sqrt{{\eps'}^2-3\varpi_\ell^2(\eps-\eps')}\Bigr.$
in order to satisfy the exact Eq.~\eqref{eq:cond}. Again, $\eps\in\left]0,1\right[$. The streaming momenta are equal in all three spatial directions, and the total streaming Lorentz factor is defined as $\gamma=(1+3\varpi_\ell^2)^{1/2}$, therefore referring to the ``left-handed'' normalized streaming momentum $\varpi_\ell$. This quantity will be used in the lower panel of Fig.~4.

\begin{figure}
\begin{center}
\includegraphics[width=95mm]{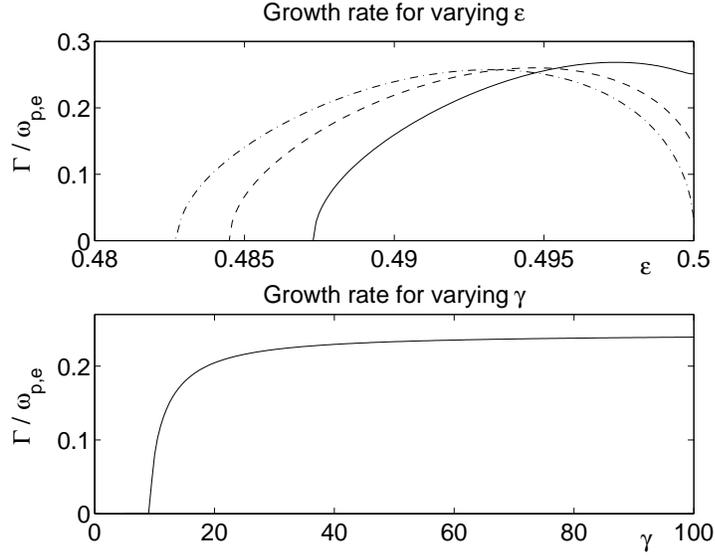}
\end{center}
\caption{The normalized growth rate $\Ga/\omega_{p,e}$ calculated for the distribution function from Eq.~\eqref{eq:beam}. The upper panel shows the growth rate as a function of $\eps$ for several cases of fixed Lorentz factors $\gamma$. The four cases are: $\gamma=10$ (solid line); $\gamma=20$ (dashed line); and $\gamma=100$ (dash-dot line). The lower panel shows the case of fixed $\eps=0.49$ for varying $\gamma$.}
\end{figure}

In Fig.~4, the growth rate for the distribution from Eq.~\eqref{eq:beam} is shown. The upper panel shows the growth rate as function of the asymmetry parameter $\eps$ using four different values of $\gamma$. Note that the growth rate decreases but remains finite as the asymmetry tends to zero (i.\,e., $\eps\rightarrow\frac{1}{2}$). The fact that the growth rate goes to zero for particular values of $\eps$ and $\gamma_\ell$ is consistent with the result of T06b that, for a two-stream distribution, a minimum streaming velocity (described by $\gamma_\ell$) is required in order to allow for unstable Weibel modes. Note that this is in contrast to the (non-relativistic) thermal distribution from Eq.~\eqref{eq:Cauchy}. The lower panel shows the growth rate as a function of the Lorentz factor $\gamma$, indicating that, for increasing $\gamma$, the growth rate finally saturates. One might expect the growth rate to decrease again, as the Lorentz factor increases further, because the relativistic mass increase makes the system stiff. The fact that the growth rate remains constant is due to a relativistic effect that, in order to satisfy Eq.~\eqref{eq:cond}, causes $\varpi_r\rightarrow\const$ for increasing $\varpi_\ell$ (i.\,e., increasing $\gamma_\ell$), as noted above. Therefore, the growth rate reaches a constant value rather than dropping towards zero.

\section{Summary and discussion}

In this paper, it was shown analytically that unstable isolated Weibel modes exist for a thermal counterstreaming plasma. These isolated modes are currently an interesting and important subject of research as they must occur in all asymmetric distributions and are, therefore, likely to be present in many astrophysical and also machine plasmas.

It was proposed that the basic situation can be represented by a non-relativistic asymmetric Cauchy distribution function, which is qualitatively comparable to a Maxwellian but has analytical advantages. Unless the asymmetry is precisely zero, isolated unstable Weibel modes are excited with growth rates of $0<\Gamma/\omega_{p,e}<0.656$, depending on the parameters and relative intensity for the two counterstreaming components that determine the asymmetry. Furthermore, the case of a relativistic two-stream distribution function was investigated, where the instability rate increases towards a maximum as the streaming Lorentz factor $\gamma$ increases from unity. Both distribution functions were constructed so that the asymmetry was tunable via an asymmetry parameter $\eps$ and, therefore, the behavior of the unstable modes could be investigated as the asymmetry tends to zero. It was shown that, unless the asymmetry is precisely zero, the growth rate remains finite. Only in the case of exact symmetry, does one have a continuum range of values for the imaginary phase velocity exists, allowing for a broad range of unstable wavenumbers. The transition between these two situations is, however, not smooth because the number of discrete modes remains constant until the asymmetry is exactly zero.

Future work should explore---both analytically and by means of self-consistent PIC simulations---under which conditions the isolated Weibel modes investigated here may develop to soliton waves, for which such isolated modes as investigated here are a necessary prerequisite.

\ack
{\it This work was partially supported the Deutsche Forschungsgemeinschaft (DFG) through grant No.~Schl201/17-1 and Sonderforschungsbereich 591.}


\end{document}